\documentclass[journal]{IEEEtran}
\usepackage{amsmath,amssymb,amsfonts}
\usepackage{subfigure}
\usepackage{graphicx}
\usepackage{caption}
\usepackage{diagbox}
\usepackage{color}
\usepackage{makecell}
\usepackage{cite}

\begin{document}

\title{Static Power Consumption Modeling and Measurement of Reconfigurable Intelligent Surfaces }

\author{Jinghe~Wang, Wankai~Tang, Shi~Jin, Xiao~Li, and~Michail Matthaiou
	
\thanks{J.  Wang, W. Tang, S. Jin and X. Li are with the National Mobile Communications Research Laboratory, Southeast University, Nanjing 210096, China  (e-mail: wangjh@seu.edu.cn; tangwk@seu.edu.cn; jinshi@seu.edu.cn; li\_xiao@seu.edu.cn).

M. Matthaiou is with the Centre for Wireless Innovation (CWI), Queen’s University Belfast, Belfast BT3 9DT, U.K. (e-mail: m.matthaiou@qub.ac.uk

The work of M. Matthaiou was supported in part by the European
Research Council (ERC) under the European Union’s Horizon 2020 Research
and Innovation Program under Agreement 101001331).

}}

\maketitle

\begin{abstract}
Reconfigurable intelligent surfaces (RISs) are anticipated to transform wireless communication in a way that is both economical and energy efficient. Revealing the practical power consumption characteristics of RISs can provide an essential toolkit for the optimal design of RIS-assisted wireless communication systems and energy efficiency performance evaluation. Based on our previous work that  modeled the dynamic power consumption of RISs, we henceforth concentrate more on static power consumption. We first divide the RIS hardware into three basic parts: the FPGA control board, the drive circuits, and the RIS unit cells. The first two parts are mainly to be investigated and the last part has been modeled as the dynamic power consumption in the previous work. In this work, the power consumption of the FPGA control board is regarded as a constant value, however, that of the drive circuit is a variant that is affected by the number of control signals and its self-power consumption characteristics. Therefore, we model the power consumption of the drive circuits of various kinds of RISs, i.e., PIN diode-/Varactor diode-/RF switch-based RIS. Finally, the measurement results and typical value of static power consumption are illustrated and discussed.
\end{abstract}


\IEEEpeerreviewmaketitle

\section{Introduction}

\IEEEPARstart{F}{uture} wireless networks must contend with extremely complex networks, expensive hardware, and increasing energy consumption as the demands for wireless network capacity grow rapidly. In recent years, both the wireless research community and industry have paid close attention to the development of RISs, which can flexibly manipulate the electromagnetic properties of radio waves.

Typically, a RIS is a 2D planar architecture that is made up of a large number of meticulously designed electromagnetic unit cells. The electromagnetic properties of these unit cells can be dynamically controlled by applying control signals, which create electromagnetic fields with controllable phase, amplitude, polarization, and frequency \cite{cui2014coding,zhang2019breaking}. Therefore, a RIS is introduced to enable the wireless propagation environment to change from passive adaptation to active control, thereby creating a smart radio environment \cite{9140329}. 

When there is not a direct link between the user equipment (UE) and the base station (BS), a RIS can be deployed to provide virtual line-of-sight (vLoS) links for signal transmission. The impinging signal is reflected by the RIS, and RIS performs 3D beamforming to improve the received signal quality by controlling the reflection phase shift of each RIS unit cell, which is expected to increase the wireless communication system's transmission rate, coverage, and energy efficiency. System performance enhancement can be achieved through joint active beamforming at the BS and passive beamforming at the RIS \cite{han2019large,wu2019intelligent,huang2019reconfigurable,guo2020weighted}. Moreover, the RIS hardware design, RIS prototyping system design, as well as experimental validation were illustrated in \cite{tang2020wireless,trichopoulos2022design}. These results are expected to pave the way for RIS deployments and practical use in the future. 

The power consumption characteristic of RIS is also an essential topic. Revealing the practical power consumption characteristics of RISs can provide an essential toolkit for the optimal design of RIS-assisted wireless communication systems and energy efficiency performance evaluation. The authors in \cite{huang2019reconfigurable} and \cite{zappone2020overhead} developed power consumption models of RIS-assisted wireless communication systems which are mainly used for solving the energy efficiency optimization problems of RISs. However, the static power consumption of the FPGA control board and the drive circuit was not considered in \cite{huang2019reconfigurable}, while the power consumption model of RIS was not illustrated in depth in \cite{zappone2020overhead}.
The authors in \cite{tang2020wireless} presented some initial measurement results on the RIS power consumption, which demonstrate that various RIS types have various power consumption characteristics. In \cite{9551980}, the power consumption of fabricated varactor-diode-based RIS was demonstrated. Nevertheless, the power consumption was not modeled and discussed thoroughly in \cite{tang2020wireless,9551980}. Against this background, we presented a precise RIS power consumption model. In \cite{wang2022reconfigurable}, a quantitative relationship between the dynamic power consumption of the PIN-diode-based RIS and the polarization mode, controllable bit resolution, and working status of RIS was proposed, while the dynamic power consumption of the varactor-diode-based RIS was described to be almost negligible. Both conclusions were validated by practical experimental results. However, static power consumption is regarded as a constant part that is simply measured via power metering sockets, which should be analyzed more accurately. Therefore, in this work, we complement the modeling of the static power consumption of RISs. Firstly, we divide the RIS hardware into three basic parts: the FPGA control board, the drive circuit, and the RIS unit cells. The first two parts are mainly investigated in the following and the last part has been modeled as the dynamic power consumption in the previous work \cite{wang2022reconfigurable}. Particularly, the power consumption model of the drive circuits is proposed, which is affected by the number of control signals and its self-power consumption characteristics. Finally, various of fabricated RISs, i.e., PIN diode-/varactor diode-/RF switch-based RIS are introduced to illustrate the practical measurement results.

\section{Static Power Consumption Modeling}

\begin{figure}
\centering
\includegraphics[height=3.5cm,width=9cm]{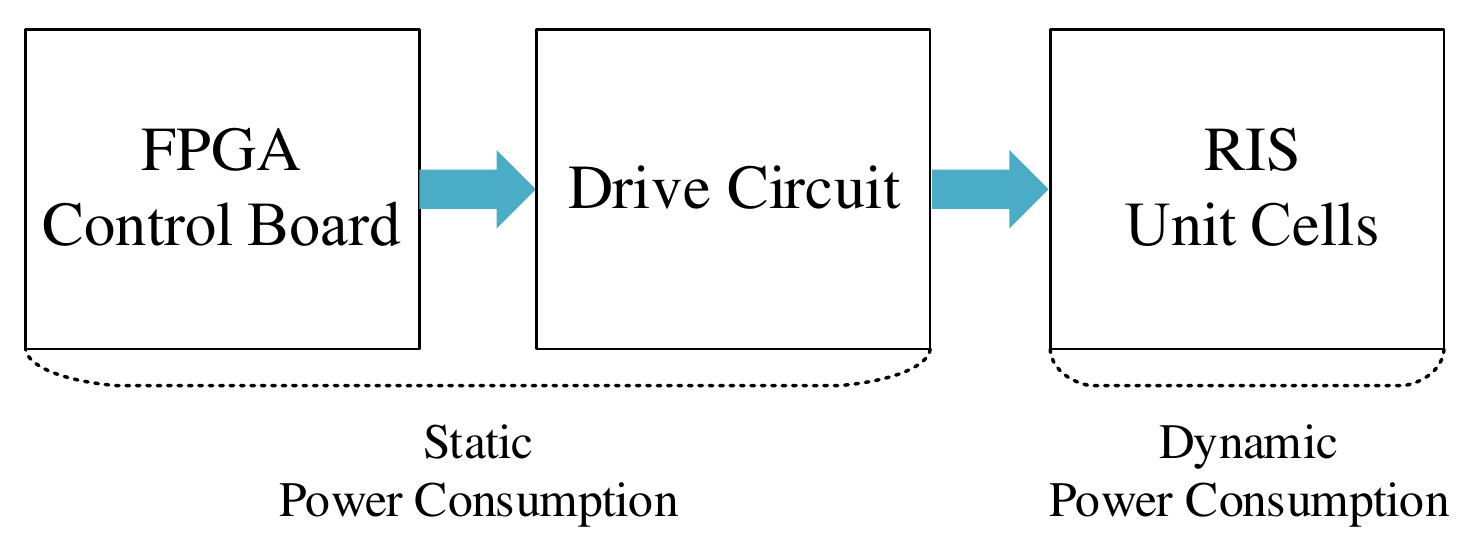}
\caption{General RIS hardware design structure can be simply divided into three parts: FPGA control board, the drive circuit, and RIS unit cells.}
\label{1}
\end{figure}

RISs often rely on adjustable electronic components, i.e., PIN diodes, varactor diodes, and RF switches to change the states of the RIS unit cells, and perform 3D beamforming to refocus the energy from the incident RF signal to a specified direction. An external controller is required to be embedded in the RIS for programming, configuration, and digital control. As shown in Fig.\ref{1}, general RIS hardware designs can be divided into three parts: the FPGA control board, the drive circuits to drive adjustable electronic components, and RIS unit cells. 

Firstly, the FPGA control board is a programmable array of logic gates, which is not only used for providing clock signal, but also for data processing, i.e., generating the corresponding RIS beamforming coding sequence configuration according to the preset algorithm based on the feedback data information. Secondly, the drive circuits are used to drive the adjustable electronic components. They are usually integrated circuits for generating variable current/voltage according to the control signals from the master FPGA  control board. Then, these adjustable electronic components are driven to achieve different coding states. At last, the RIS unit cells are artificial electromagnetic structures that are composed of substrates, metallic vias, metal patches, and adjustable electronic components. They are regularly arranged on a 2D surface and generate different 3D beamforming patterns according to the combination of different coding sequences.  

Overall, the total power consumption dissipated to operate the RIS consists of two parts, one is the static power consumption generated by the FPGA control board and drive circuits, and the other is the dynamic power consumption generated by the RIS unit cells. Therefore, the total power consumption modeling of RIS can be expressed as follows:
\begin{equation}
P_{\mathrm{RIS}}=P_{\text {static }}+P_{\text {dynamic, }}
\end{equation}
where $P_{\text {static }}$ is the static power consumption and $P_{\text {dynamic }}$ is the dynamic power consumption. In \cite{wang2022reconfigurable}, $P_{\text {dynamic }}$ has been precisely modeled. More specifically, for a PIN-diode-based RIS, $P_{\text {dynamic }}$ is related to the number of RIS unit cells, the polarization mode, bit resolution, and the coding state of RIS unit cells. For the varactor-diode-based RIS, $P_{\text {dynamic }}$ is almost zero, since the current in the varactor can be negligible when it works, even though the RIS unit cells are large in number.

However, $P_{\text {static }}$ has not been thoroughly investigated in \cite{wang2022reconfigurable}. According to whether the RIS hardware power consumption is affected by the number of control signals, the static power consumption dissipated to operate the RIS is the superposition of two parts, one is the power consumption of the FPGA control board $P_{\text {control board}}$, and the other is the power consumption of drive circuits $P_{\text {total drive circuits }}$. The static power consumption modeling of RIS can be expressed as follows:

\begin{equation}
P_{\mathrm{static}}=P_{\text {control board}}+P_{\text {total drive circuits. }}
\end{equation}
Generally, an FPGA selected as the master control board needs to have enough arithmetic power to cover all data processing, thus $P_{\text {control board}}$ can be regarded as a constant value. However, $P_{\text {total drive circuits }}$ is variable that is related to the types of adjustable electronic components, the number of control signals, and self-power consumption characteristics.

\begin{figure*}[!t]
\large
\begin{equation}
P_{\mathrm{static}}=\left\{\begin{array}{cc}
P_{\text {control board}}+\left\lceil\frac{\sum_{i=1}^N B_i}{N_{\text {g}} \cdot N_{\mathrm{s}}} \right\rceil \cdot P_{\text {drive circuit }}, & \text { PIN-diode/ RF switch-based RIS }  \vspace{0.3cm} \\
P_{\text {control board}}+\left\lceil\frac{N}{N_{\text {g}} \cdot N_{\mathrm{s}}} \right\rceil \cdot P_{\text {drive circuit }}, & \text { \quad varactor-diode-based RIS } 
\end{array}\right.
\end{equation}
\hrulefill
\vspace{-0.3cm}
\end{figure*}
\begin{itemize}
\item $P_{\text {total drive circuits }}$ is related to the types of adjustable electronic components since different components utilize different types of drive circuits. For example, PIN-diode-based RIS unit cells can be driven by shift registers; varactor-diode-based RIS unit cells can be driven by digital-to-analog converters (DACs) and operational amplifiers (op-amps), PWM signals + level regulator \cite{9551980}, or CMOS logic circuits; RF switch-based RIS unit cells can be driven by FPGAs and shift registers.
\item $P_{\text {total drive circuits }}$ is related to the number of control signals. Specifically, it is related to the number of adjustable electronic components $N_{\text {c}}$ and the control Degree-Of-Freedom (DoF) of RIS (i.e., unit cell control, row control, column control, and sub-array control). RIS unit cells divided into the same group can use the same control signals, and $N_{\text {g}}$ is denoted as the number of RIS unit cells that are in the same group with the same control signal.
\item Total drive circuits are composed of multiple individual drive circuits. Therefore, $P_{\text {total drive circuits }}$ is related to the self-power consumption characteristics of the individual drive circuit. $N_{\text {s}}$ is denoted as the number of control signals generated by each drive circuit and $P_{\text {drive circuit }}$ is denoted as the rated power consumption of a single drive circuit.
\end{itemize}
Based on the above, $P_{\text {total drive circuits }}$ can be expanded as follows:

\begin{equation}
\begin{aligned}
P_{\text {total drive circuits }} & =N_{\text {drive circuit }} \cdot P_{\text {drive circuit }} \\
&=\left\lceil\frac{N_{\mathrm{c}}}{N_{\text {g}} \cdot N_{\mathrm{s}}} \right\rceil \cdot P_{\text {drive circuit }},
\end{aligned}
\end{equation}
where $\lceil\cdot\rceil$ is the round-up symbol, $N_{\mathrm{c}}$ is the number of required adjustable electronic components, $N_{\text {g}}$ is the number of RIS unit cells that are in the same group with the same control signal, and $N_{\mathrm{s}}$ is the number of control signals generated by a single drive circuit.
For PIN-diode-based RIS, $N_{\mathrm{c}}$ can be expressed by 
\begin{equation}
N_{\mathrm{c}}=\sum_{i=1}^N B_i,
\end{equation}
where $N$ is the number of RIS unit cells. Specifically, each PIN diode requires an input control signal to switch between 1 bit-binary state. Thus, $B_i$ control signals are required for the ${i}$-th unit cell with $B_i$ bits. For varactor-diode-based RIS, a single varactor can be coded to different states by applying different magnitudes of reverse bias voltage to it, thus $N_{\mathrm{c}}=N$. For RF switch-based RIS, the ${i}$-th RF switch requires $B_i$ control signals for $B_i$-bit coding states.
therefore, $N_{\mathrm{c}}$ is the same as PIN-diode-based RISs in (5). In conclusion, $P_{\text {static }}$ is summarized as (3) at the top of this page. Equations (1), (3) in this work, and  equation (2) in \cite{wang2022reconfigurable} form the full RIS power consumption modeling together.

\section{Practical Static Power Consumption Analysis}
In the following section, we provide a detailed description and measurement results of the static power consumption of  practical RISs. Overall, the FPGA control board is like the “brain” of the RIS, which needs to generate the coding sequences for regulating the RIS. A flow diagram for realizing a RIS triggered by the FPGA hardware is presented in Figure 7i in \cite{cui2014coding}. In the measured PIN-diode-based RIS and RF switch-based RIS, the commercial-level FPGA XC7K70T is embedded as the master FPGA control board. With a working voltage of 24 V, the current of it is measured as 0.2 A. Therefore, $P_{\text {control board}}=4.8$ W. Some entry-level FPGAs, i.e., development board with Xilinx ZYNQ7100 are sufficient for RIS prototypes with fewer RIS unit cells or low data processing speed requirements, which consumes only 1.5 W like the RIS prototype in \cite{9551980}.

\subsection{PIN-diode-based RIS}

\begin{figure}
\centering
\includegraphics[height=8cm,width=8cm]{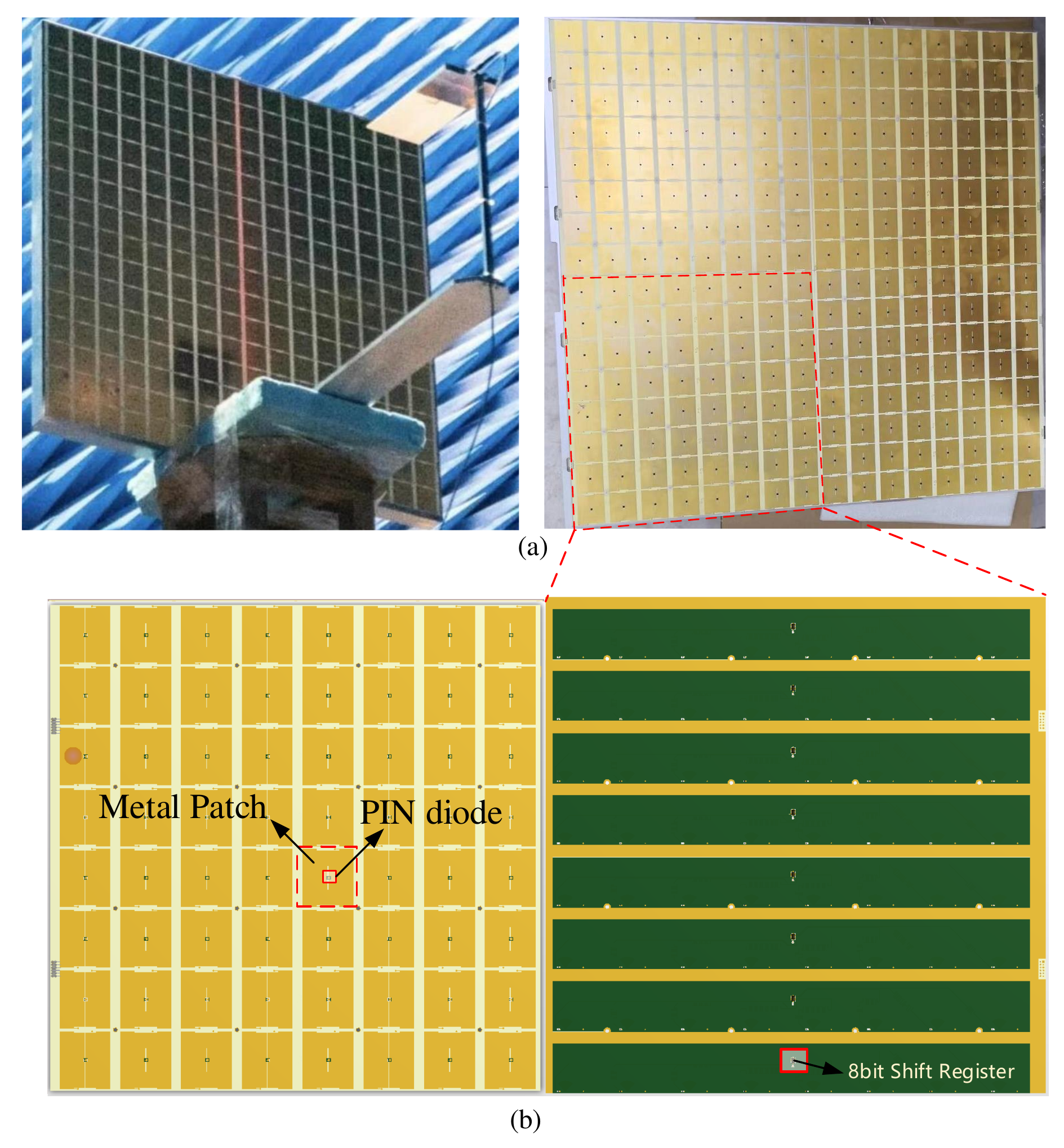}
\caption{Photograph of the fabricated PIN-diode-based RIS. (a) A complete $16 \times 16$ RIS. (b) An $8 \times 8$ sub-RIS structure.}
\label{2} 
\end{figure}

\begin{figure}
\centering
\includegraphics[height=3.5cm,width=9cm]{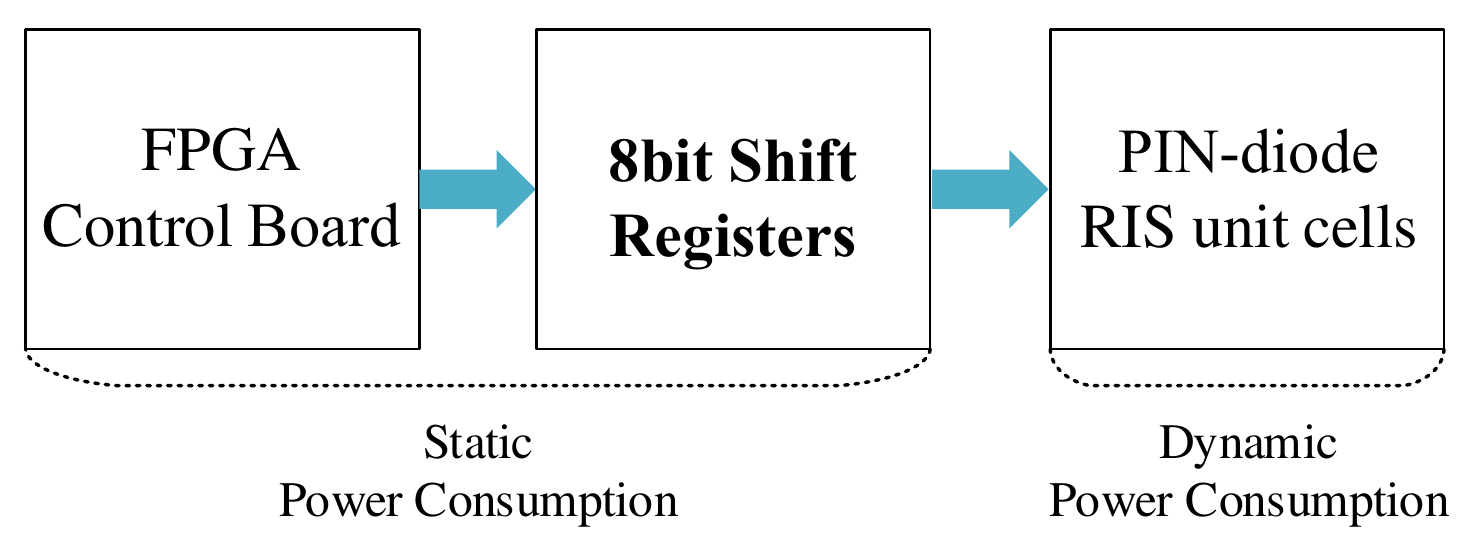}
\caption{The fabricated PIN-diode-based RIS hardware design structure.}
\vspace{-0.2cm}
\label{3}
\end{figure}

The RIS utilized for the first measurement belongs to the family of PIN-diode-based RIS, which is phase-programmable with 1-bit coding, as shown in Fig. \ref{2}. It can control the phase shift of reflected EM waves at the operating frequency $f=3.5$ GHz by unit cell independently. Four $8 \times 8$ sub-RISs form a complete $16 \times 16$ RIS. The PIN-diode drive circuits are essential since they affect the speed of the PIN diodes switching between “on/off” (coding states). By applying PIN-diode drive circuits, control signals from the FPGA master control board can be changed into output signals with current-driven capability. In the measured PIN-diode-based RIS, the SN74LV595A 8-bit shift registers are applied as drive circuits, as shown in Fig. \ref{3}. 

Specifically, an 8-bit shift register is embedded for 8 unit cells on a column in an $8 \times 8$ sub-RIS, which can turn the serial input into multiple parallel outputs. Therefore, $8 \times 4=32$ shift registers are required in the complete $16 \times 16$ RIS. When 8-bit shift registers are working at the voltage $\mathrm{V}_\text {cc} =$ 3.3 V, the current we measured is $\mathrm{I}_\text {cc}  = 20 $ $\mu$A, thus, $P_{\text {drive circuit }} = 0.066$ mW. The key parameters of this RIS in the model are $P_{\text {control board }} = 4.8$ W, $N_{\mathrm{c}}=256$, $N_{\mathrm{g}}=1$, $N_{\mathrm{s}}=8$, therefore $N_{\text {drive circuit}}= 32$, and $P_{\text {total drive circuits }}= 32 \times P_{\text {drive circuit }}= 2.112 $ mW. As illustrated in the measurement results, it can be seen that applying shift registers as drive circuits is an energy-efficient solution, ensuring that $P_{\text {total drive circuits }}$ is about mW-level for hundreds/thousands of unit cells.


\subsection{Varactor-diode-based RIS}

\begin{figure}
\centering
\includegraphics[height=7cm,width=9cm]{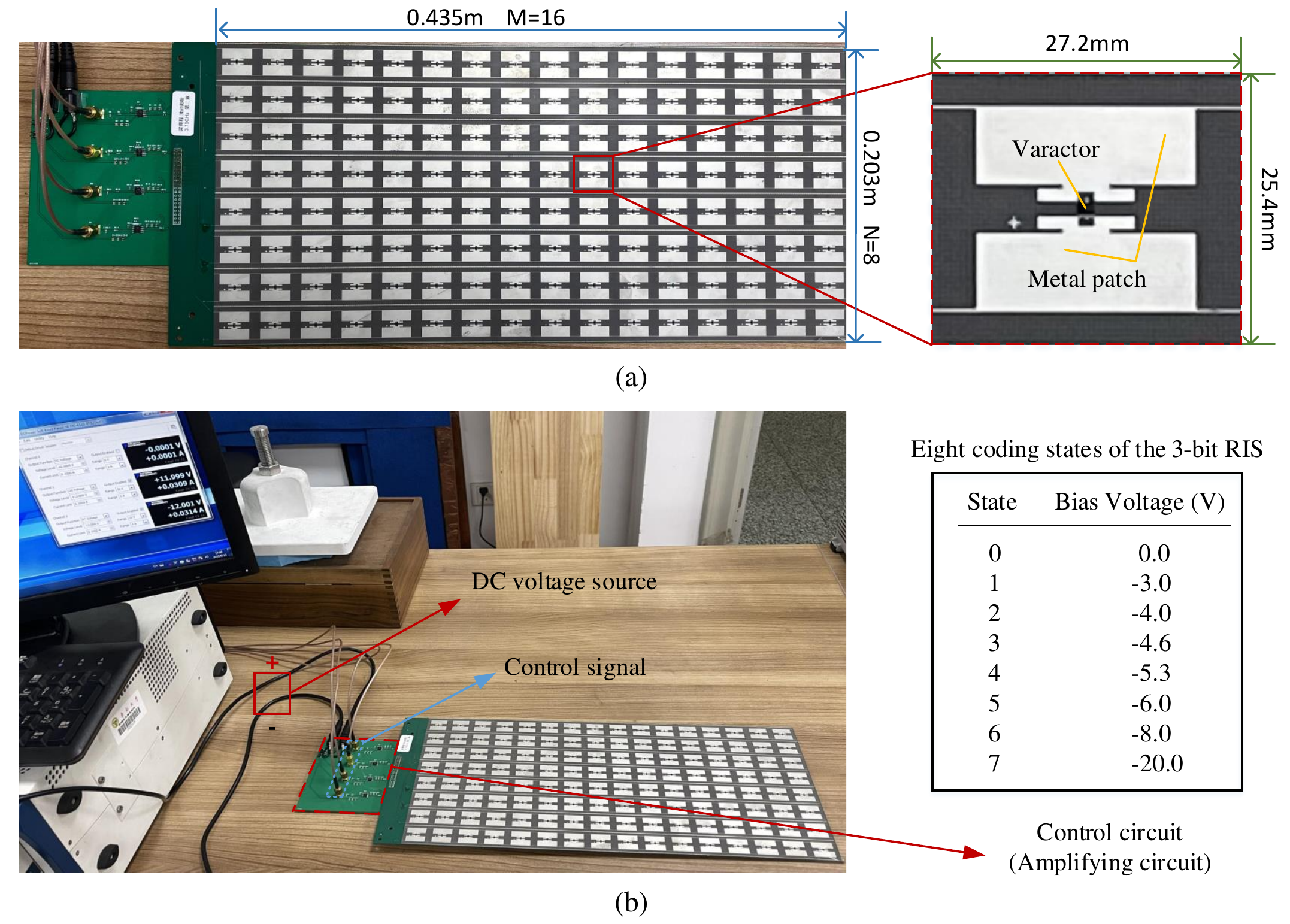}
\caption{Photograph of the fabricated varactor-diode-based RIS. (a) A complete RIS and its unit cell structure. (b) Practical measurement and a typical example of 8 coding states of the RIS.}
\label{4} 
\vspace{-0.2cm}
\end{figure}

\begin{figure}
\centering
\includegraphics[height=3.5cm,width=9cm]{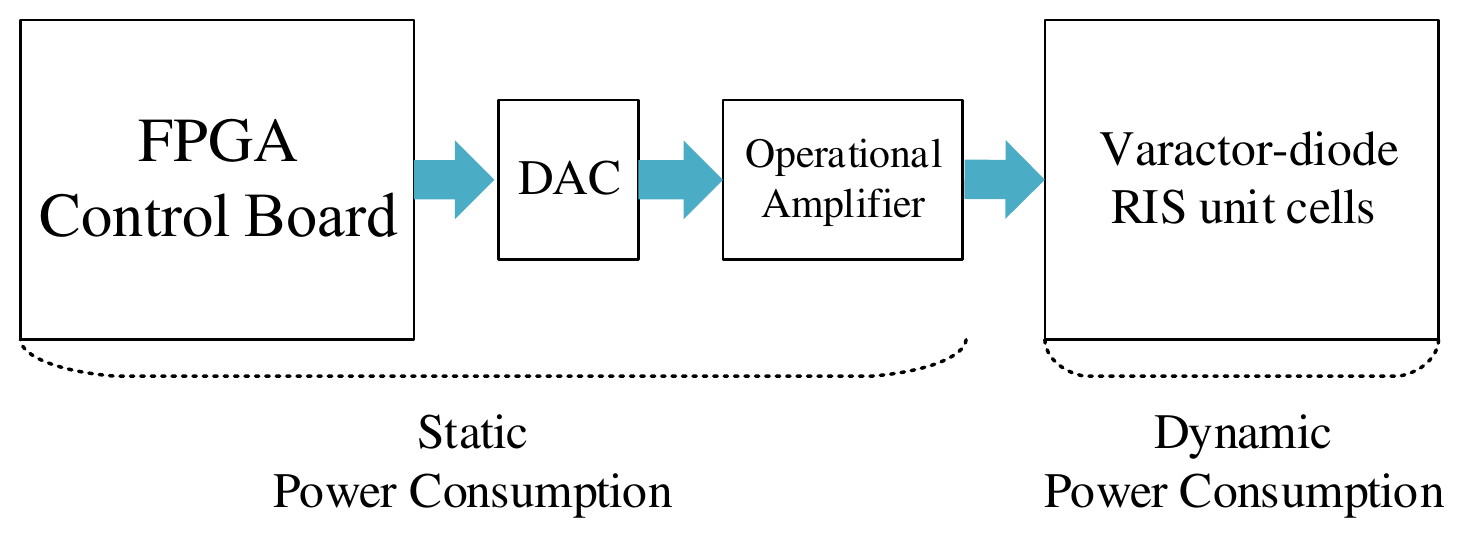}
\caption{The fabricated varactor-diode-based RIS hardware design structure.}
\label{5}
\vspace{-0.2cm}
\end{figure}

The RIS utilized for the second measurement belongs to the family of varactor-diode-based RIS, as shown in Fig. \ref{4}. It is  operated at $f=3.2 $ GHz by column control \cite{liang9632392}. The RIS is composed of $N=8$ columns with $M=16$ unit cells in each column. The varactor diode is embedded to bridge the metal patches and operated as an adjustable device and is operated under reverse bias voltages. The typical 8 coding states corresponding to the bias voltages applied on the varactors are illustrated in Fig. \ref{4}(b). Varactor-diode-based RIS unit cells can be driven by DAC + op-amp, PWM signals + level regulator, or CMOS logic circuits. In the measured varactor-diode-based RIS, the DAC and op-amp are applied, as shown in Fig. \ref{5}. DACs take digital inputs and generate analog outputs that provide bias voltages of varactor diodes. However, the bias voltage generated by the common DAC is usually low and can not achieve a high one like -20 V in Fig. \ref{4}(b). Therefore, the op-amp is necessary for the operation amplification function. Moreover, the DAC3484 is applied as a DAC, and the AD8021 low-noise, high-speed amplifier is applied as an op-amp. 

In our measurement, 2 columns of RIS unit cells are in the same group with the same control signal, thus $N_{\text {g}}= 16 \times 2 =32$. Specifically, DAC3484 consumes about 250 mW for generating one signal. When the AD8021 works at the voltage $\mathrm{V}_\text {cc}= \pm 12 $ V, the current we measured is 7.5 mA, so the power consumption of the AD8021 is 180 mW. Thus, $P_{\text {drive circuit }}= 250 + 180= 430 $ mW. The key parameters of this RIS in the model are $N_{\mathrm{c}}=128$, $N_{\mathrm{g}}=32$, and $N_{\mathrm{s}}=1$, therefore $N_{\text {drive circuit}}= 4$, and $P_{\text {total drive circuits }}= 4 \times P_{\text {drive circuit }} = 1720 $ mW. 


\subsection{RF Switch-based RIS}

\begin{figure}
\centering
\includegraphics[height=8.5cm,width=8cm]{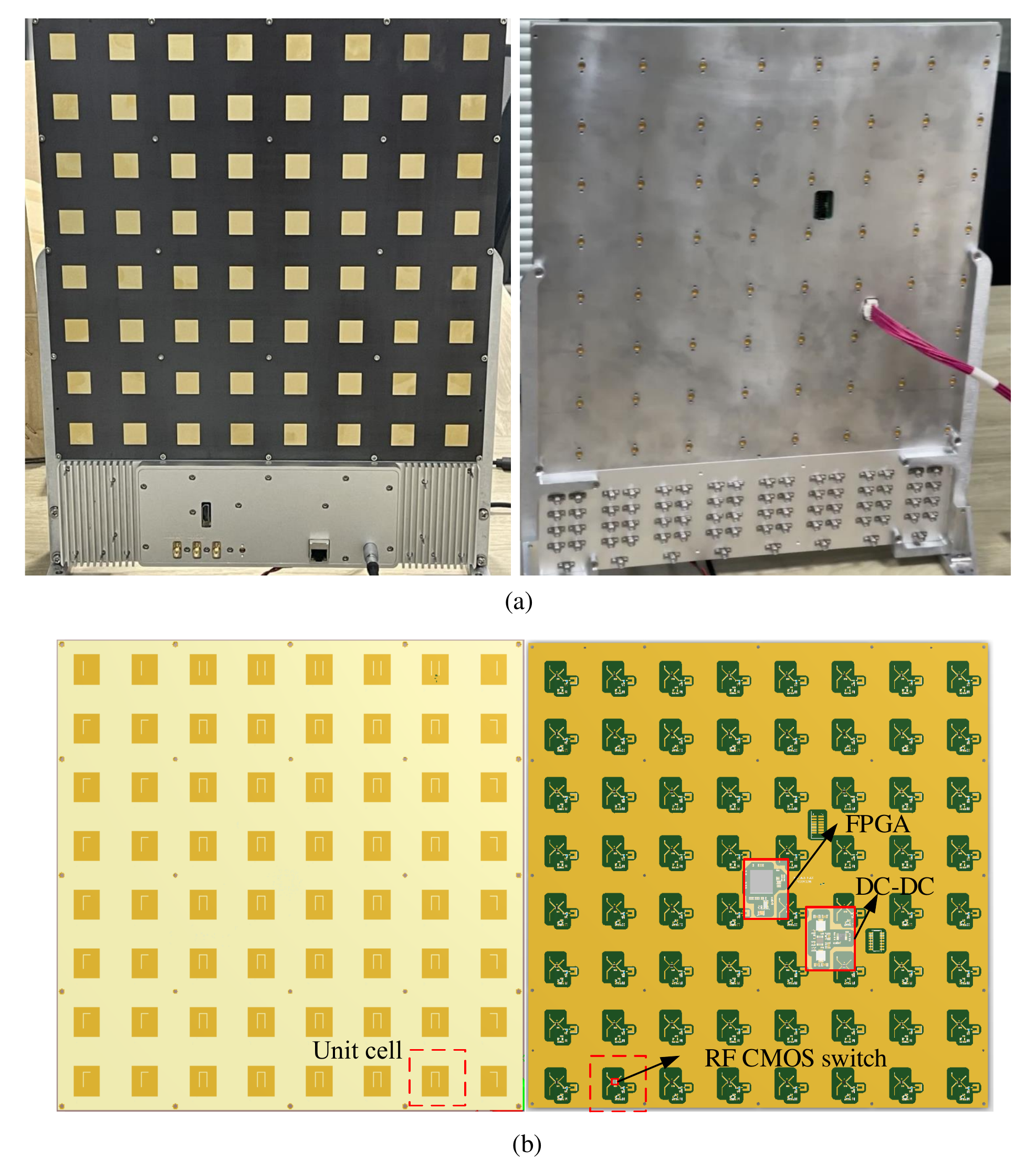}
\caption{Photograph of the fabricated RF switch-based RIS. (a) A complete RIS physical diagram. (b). $8 \times 8$ RIS structure and parameters' illustration.}
\label{6}  
\vspace{-0.2cm}
\end{figure}

\begin{figure}
\centering
\includegraphics[height=3.5cm,width=9cm]{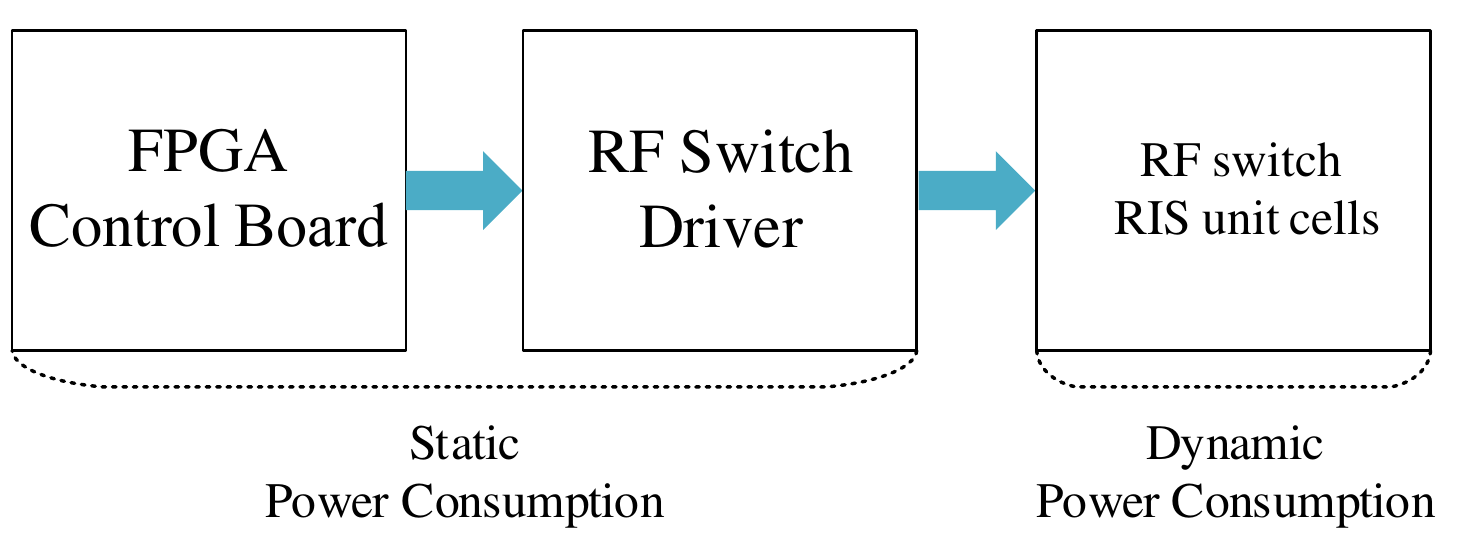}
\caption{The fabricated RF switch-based RIS hardware design.}
\label{7}
\vspace{-0.2cm}
\end{figure}

Common solutions implement PIN-diode-based RISs with a finite-bit resolution phase shift and implement varactor-diode-based RISs with both discrete and continuous control of the phase shift. Recently, RF switch-based RISs have emerged since their implementations entail low costs \cite{rossanese2022designing}. Not only that, in addition to producing phase shifts onto impinging signals in a programmable manner, an RF switch-based RIS can configure individual cells to fully absorb the energy of RF signals, which provides open several opportunities. 
The RIS utilized for the third measurement belongs to the family of RF switch-based RIS, as shown in Fig. \ref{6} and it is composed of $8 \times 8=64$ RIS unit cells, with each unit cell being regulated independently. A RIS unit cell consists of an RF CMOS switch embedded in the metal patch. For an RF switch-based RIS, RF switch drivers are needed to provide logic levels to control the RF switches switching between different coding states, as shown in Fig. \ref{7}. 

In the measured RF switch-based RIS, FPGA XC3S400AN is embedded as an RF switch driver to control 64 RIS unit cells. When FPGA XC3S400AN is working at the voltage $\mathrm{V}_\text {cc} =$ 12 V, the current is measured as 20 mA. Therefore, $P_{\text {drive circuit}} = 240 $ mW.  The key parameters of this RIS in the model are $N_{\mathrm{c}}=64$, $N_{\mathrm{g}}=1$, and $N_{\mathrm{s}}=75$ (a single FPGA XC3S400AN can control about 75 RF switches), therefore $N_{\text {drive circuit}}= 1$, and $P_{\text {total drive circuits }}= P_{\text {drive circuit }} = 240$ mW.

\subsection{Discussion}
The static power consumption of various RISs is now summarized, and the results are all supported by the practical measurements and the product data sheets. First of all, the FPGA control board consumes Watt-level power for data processing.
Secondly, even though the $P_{\text {dynamic}}$ of varactor-diode-based RIS is almost zero, the drive circuits of it are often much more complicated and energy-hungry, like $P_{\text {drive circuit }}= 430 $ mW for each control signal with continuous bias voltage variation. Moreover, there still remain challenges for designing and manufacturing the varactor-diode-based RIS in high-frequency bands (e.g., mmWave band). Oppositely, the drive circuits of PIN-diode-based RIS are energy-efficient, which are mWatt-level consuming, like $P_{\text {drive circuit }}= 0.066 $ mW for each 8-bit shift register and $ 0.066/8= 8 \times 10^{-3} $ mW for each control signal. However, the $P_{\text {dynamic}}$ of PIN-diode-based RIS can be very large. In \cite{wang2022reconfigurable}, 
the $P_{\text {PIN}}$ of each unit cell in $1^{\#}$ RIS is measured as 12.6 mW, and the maximum $P_{\text {dynamic}}$ is about $90$ W when all unit cells are encoded as “1”. As a matter of fact, discrete-type varactor-diode-based RISs are promising for energy-efficient design, since its $P_{\text {dynamic}}$ is almost zero and its drive circuit $P_{\text {total drive circuits }}$ may sharply decline for only generating discrete-level bias voltage.

In addition, the $P_{\text {total drive circuits}}$ of RF switch-based RISs is measured as 240 mW for a single drive circuit (FPGA) to control 64 unit cells. Actually, $P_{\text {total drive circuits}}$ can be further reduced for the low power consumption design. On the one hand, the master FPGA control board can directly provide control signals for RF switch-based RIS unit cells, in which way the drive circuits can be omitted, thus $P_{\text {total drive circuits}} = 0$. However, when a huge number of RIS unit cells are embedded, this approach is invalid since the master FPGA control board can not provide control signals to unit cells simultaneously. On the other hand, shift registers can also be utilized as RF switch drivers, in which the $P_{\text {total drive circuits}}$ of RF switch-based RISs is comparable to that of PIN-diode-based RISs. Additionally, the $P_{\text {dynamic}}$ of RF switch-based RISs is usually ${\mu}$Watt-level power-consuming, about $3.3$ V $\times 150 \mu$A $= 495 \mu$W for each unit cell.
Overall, the low power consumption design of RISs is an open research area for further investigation.


\section{Conclusion}
In this work, we divided the RIS hardware into three basic parts: the FPGA control board, the drive circuits, and the RIS unit cells. The previous work has modeled the dynamic power consumption of RIS unit cells, thus we investigated the static power consumption of the FPGA control board $P_{\text {control board}}$ and that of the drive circuits $P_{\text {total drive circuits }}$. 
We articulated that the $P_{\text {control board}}$ can be regarded as a constant value and $P_{\text {total drive circuits }}$ is affected by the number of control signals and its self-power consumption characteristics. Particularly, $P_{\text {total drive circuits }}$ is modeled for various kinds of RIS, i.e., PIN diode-/Varactor diode-/RF switch-based RIS, and measurement results validated the model of $P_{\text {static}}$. Overall, the dynamic power consumption modeling $P_{\text {dynamic}}$ in the previous work in \cite{wang2022reconfigurable} and the static power consumption modeling $P_{\text {static}}$ in this work form the full RIS power consumption modeling together. 

\vspace{-0.1cm}

\ifCLASSOPTIONcaptionsoff
  \newpage
\fi

 \bibliographystyle{IEEEtran}
 \bibliography{mybib}
 
\end{document}